\documentclass[twocolumn,superscriptaddress,showpacs,aps,prb]{revtex4}
\usepackage{mathtext}
\usepackage{graphics}
\usepackage[dvips]{epsfig}
\usepackage{amsmath}
\usepackage{amsmath}

\def\XXint#1#2#3{{\setbox0=\hbox{$#1{#2#3}{\int}$}
\vcenter{\hbox{$#2#3$}}\kern-.55\wd0}}

\newcounter{fig}

\begin{document}
\title{ Effect of electron-phonon interactions on Raman line at ferromagnetic ordering}
\author{L.A. Falkovsky}
\affiliation{ Landau Institute for Theoretical Physics, Chernogolovka 142432}
\affiliation{Verechagin Institute of the High Pressure
Physics, Troitsk 142190}

 \begin{abstract}
 
The theory of Raman scattering in half-metals by optical phonons interacting with conduction electrons   is developed. We evaluate  the effect of electron-phonon interactions  at ferromagnetic ordering  in terms of the Boltzmann equation for carriers. The chemical potential is found to decrease with temperature decreasing.   Both the linewidth and frequency shift exhibit a dependence on temperature.

\pacs{42.50.Nn 63.20.-e 75.30.Ds 78.30.-j}
\end{abstract}

\maketitle

\section{Introduction}

Recently, the Raman scattering in the half-metallic CoS$_2$  was studied \cite{La} in the wide temperature region. The $\omega=400 $ cm$^{-1}$ Raman line, observed  previously  at room  temperature in Ref. \cite{AP,ZST}, demonstrates a particular behavior nearby the ferromagnetic transition  at $T_c=122$ K. The unusual large Raman linewidth and shift of the order of 10 cm$^{-1}$ were observed. 
The reflectivity singularities of  CoS$_2$ were explained in Ref. \cite{YMM} by 
the temperature variation of the electronic structure.  Another example of the electron-phonon interactions is given in  Ref. \cite{LM} in order to explain the phonon singularity at the $\Gamma$ point in graphene.
 The  electron-phonon interactions should be considered as well in the interpretation of the observed Raman scattering around the Curie temperature. 

Thermal broadening of  phonon lines in the Raman scattering is usually described in terms of three-phonon anharmonicity, i.e. by the decay of an optical phonon with a frequency $\omega $ in two phonons.
The simplest case when the final state has two acoustic phonon from one branch (the Klemens channel) was theoretically studied  by Klemens \cite{Kl},
who obtained the temperature dependence of the Raman linewidth. The corresponding lineshift was  considered in Refs. \cite{BWH, MC}.  
This theory  was compared in works \cite{BWH,MC,DBM}  with experimental data for Si, Ge, C, $\alpha-$Sn.
 A model was also considered with the phonons in the final state from different branches. It was found that  anharmonic interactions of the forth order should   be disregarded at high temperatures $T> 300$ K. 

The situation is more complicated in substances with magnetic ordering. The interaction of phonons with magnons in antiferromagnets was discussed  in the review article \cite{GZ} and more recently  in the analysis of the thermal conductivity \cite{MOZ},  the spin Seebeck effect \cite{UTH,JYM},  high-temperature superconductivity \cite{NMH}, and optical spectra \cite{KKP}.
The magnon-phonon interaction   results in the magnon damping  \cite{Wo}, however, no effect for phonons was observed.  The influence of antiferromagnetic ordering is considered in Ref. \cite{DPV}, where   the line shift was only calculated. Damping of the optical phonons was found \cite{MPB} to become large  in the rare-earth Gd and Tb below the Curie temperature achieving a value of 15 cm$^{-1}$, which is much greater than the   three-phonon interaction effect.

Despite attracting considerable interest for half a century since
the pioneering work by Fr\"ohlich, the problem of electron-phonon
interaction is still far from being solved. Migdal \cite{Mi}
developed a consistent many-body approach based on the Fr\"ohlich
Hamiltonian for interaction of electrons with acoustic (sound) phonons. As
Migdal showed ("the Migdal theorem"), the vertex corrections for
acoustic phonons are small by the  adiabatic parameter
$\sqrt{m/M}$, where $m$ and $M$ are the electron and ion masses,
respectively. The theory described correctly the electronic
lifetime, renormalization of the Fermi velocity $v_F$ and acoustic
phonon attenuation but resulted in a strong renormalization of the
 sound velocity $\tilde{s}=s(1-2\lambda)^{1/2}$, where $\lambda$
is the dimensionless coupling constant. For sufficiently strong
electron-phonon coupling $\lambda \to 1/2$, the phonon frequency
approached to zero marking an instability point of the system.
Instead, one would
intuitively expect the phonon renormalization to be weak along
with the adiabatic parameter.

This discrepancy was resolved by Brovman and Kagan \cite{BK}
almost a decade later (see also \cite{G}).
They demonstrated the shortcomings of the Fr\"ohlich model that
gave an anomalously  large phonon renormalization.
Employing the Born--Oppenheimer
(adiabatic) approximation (see, e.g., \cite{BHK}),
they found that there are two terms in the second order
perturbation theory, which compensate each other
making a result
small by the adiabatic parameter. Namely, when calculating the
phonon self-energy function $\Pi (\omega,k)$ with help of the
diagram technique, one should eliminate an  adiabatic
contribution of the Fr\"ohlich model by subtracting $\Pi
(\omega,k)-\Pi(0,k)$.

The interaction of electrons with optical phonons was first
considered by Engelsberg and Schrieffer \cite{ES} within Migdal's
many-body approach for dispersionless phonons. They predicted a
splitting of the optical phonon  at finite wavenumbers $k$ into
two branches. Ipatova and Subashiev \cite{IS} calculated later on
the optical phonon attenuation in the collisionless limit and
pointed out that the Brovman-Kagan renormalization should be carried
out for optical phonons  in order to obtain correct phonon
renormalization.
 In the  paper
\cite{AS}, Alexandrov and Schrieffer corrected the calculational
error of Ref.\
\cite{ES} and argued that no splitting was found in fact. Instead,
they predicted an extremely strong dispersion of optical phonons,
 $\omega_k=\omega_0+\lambda v_F^2 k^2/3\omega_0$,
due to the coupling to electrons. The large phonon dispersion is
a typical result of Migdal's theory \cite{AGD} using the Fr\"olich
Hamiltonian. No such a dispersion has ever been observed
experimentally. The usual dispersion of optical phonons
in metals has the order of the sound velocity.
 Reizer \cite{R}  stressed the
importance of screening effects which should be taken into account.
The works \cite{AS,R} are limited to the case of
collisionless both electron and phonon systems. Moreover, only
the phonon renormalization was considered with
no results available for the attenuation of optical phonons.

A different from many-body technique  semiclassical approach
 based on the Boltzmann equation and the equations of the
theory of elasticity was developed in the papers by Akhiezer,
Silin, Gurevich, Kontorovich, and many others (we refer the reader
to the review \cite{Kon}). This approach
was compared with various experiments, such as
attenuation of sound waves, effects of
strong magnetic fields,   crystal anisotropy, and sample surfaces
on the sound attenuation, and so
on. It can be applied to the problem of the
electron--optical-phonon interaction  \cite{MF} as well.

 In the
previous paper \cite{Fal}, we have developed a quantum theory for the optical phonon  attenuation and
 shift induced by  the interband electron transitions and tuned with a temperature variation.
Now we consider the optical phonon renormalization as a result of the electron-phonon interaction taking into account  ferro-magnetic ordering. We argue that the reasonable phonon  damping and shift can be obtained using the semiclassical  Boltzmann equation for electrons and the motion equation of phonons coupled by the deformation potential.

\section{Electron-phonon interactions at  ferromagnetic ordering }

  \begin{figure}[]
\resizebox{.5\textwidth}{!}{\includegraphics{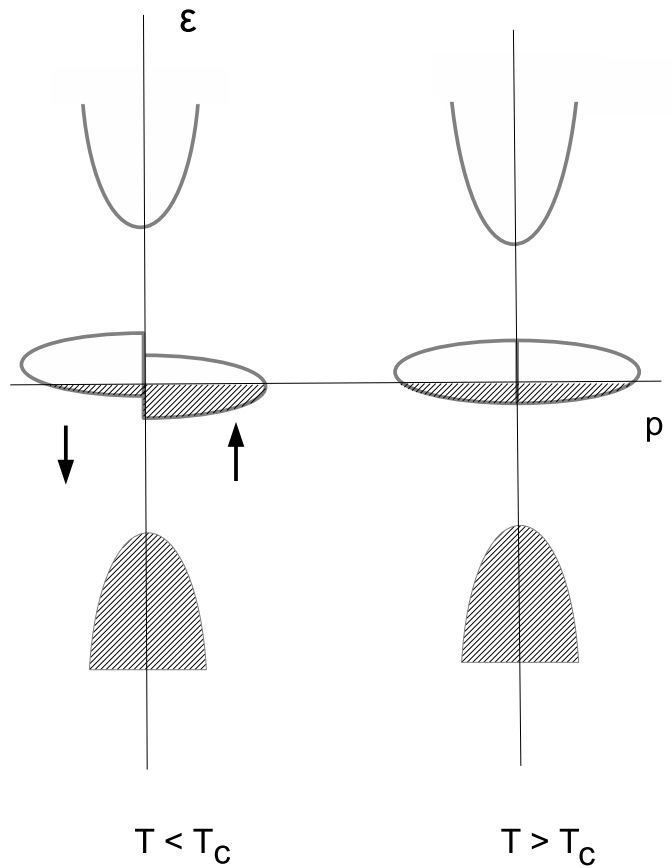}}
\caption{(Color online)  Proposed band scheme for two electron spin projections.}
\label{bands}
\end{figure}
We assume that the electron bands in CoS$_2$ have a form shown in Fig. \ref{bands}. 
The  ferromagnetic ordering results in the spin  splitting   $\mu H_e$ of the unfilled half-metallic band 
\begin{equation}
\begin{array}{c}
\displaystyle{\varepsilon_{\uparrow}({\bf p})=\frac{p^2}{2m^{\vee}}}-\mu H_e\quad \text {and} \quad
\displaystyle{\varepsilon_{\downarrow}({\bf p})=\frac{p^2}{2m^{\vee}}}+\mu H_e\,
 \end{array}
 \label{sp}\end{equation}
 in the effective Weiss field  $ H_e$.  
   While the temperature decreases, the magnetization,  determined in the mean field approximation as 
\begin{equation}M=M_0\sqrt{1-(T/T_c)^2}\,,\label{mag}\end{equation}
appears   according to experimental data  in CoS$_2$ at  $T_c=122$ K approximately, and the spin splitting is proportional to the magnetization. 

We write the interaction of electrons with the optical phonon $u_i$ as the deformation potential
\begin{equation}H_{int}=\frac{u_i}{N}\sum_{s}\int \frac{d^3{\bf p}}{(2\pi\hbar)^3}  \zeta_i({\bf p})f({\bf p})\,,\label{in}\end{equation}
where $N\sim 1/a^3$ is a number of cells in the volume unit and $a$ is the interatomic distance. For the acoustic phonon -- electron  interaction, we should  substitute
the strain tensor $u_{ij}$ instead of the displacement $u_i$ in order to satisfy the translation symmetry of the lattice.

\begin{figure}[]
\resizebox{.4\textwidth}{!}{\includegraphics{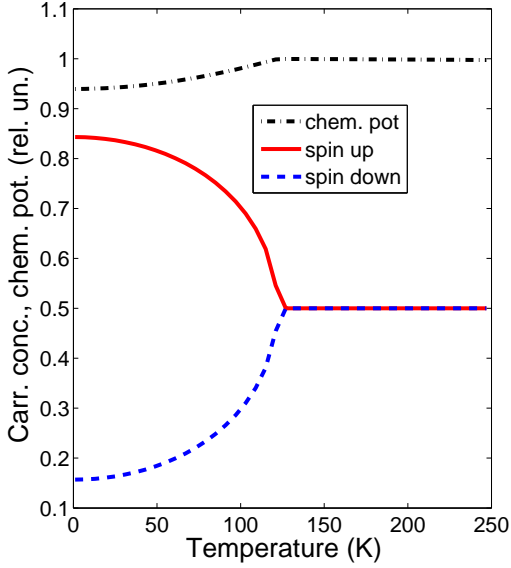}}
\caption{(Color online) Calculated temperature dependence of the carrier concentration  for the spin up and spin down (relative to the total concentration at temperatures above the temperature of ferromagnetic ordering), and the dependence of the chemical potential (dash-dotted line).}
\label{rho}
\end{figure}

The Boltzmann equation for the nonequilibrium part of the distribution function $f({\bf p})$ has the form
\begin{equation}[-i(\omega-{\bf k\cdot v})+\tau^{-1}] f({\bf p})=-\frac{\partial f_0 }{\partial \varepsilon}[e{\bf v\cdot E}-i\omega u_i\zeta_i({\bf p})]\,,\label{df}\end{equation}
where $f_0$ is the equilibrium distribution function.
We omit in the Boltzmann equation (\ref{df}) the spin index $s$, which determines all the electron parameters. The electron collision frequency $\tau^{-1}$ takes into account  the collisions with impurities and  phonons. The collision frequency is calculated for CoS$_2$  in the Debye model with the temperature $T_D= 500$ K. 
One can see from Eq.(\ref{df}) that 
the condition
\[<\zeta_i>=0\,,\]
 have to be satisfied in order to  obey the  current continuity equation,
where the  brackets mean  averaging over the Fermi surface  for temperatures $T\ll \varepsilon_F$.

In the ferromagnetic phase while the temperature changes, the carriers overflow from one spin state in another, but the total number of carriers
\begin{equation}
N=\sum_s\int\frac{d^3{\bf p}}{(2\pi\hbar)^3}f_0(\varepsilon_s)
\label{num}\end{equation}
 remains to be constant. This condition determines the chemical potential and the carrier concentration with the spin up and spin down, shown in Fig.\ref{rho}. All  figures correspond here and what follows to  the carrier concentration $N=10^{21}$ cm$^{-3}$ in the considered band with the  Fermi energy $\mu=0.36$ eV  above the Curie temperature.  

Let us write the  motion equation for the phonon mode in  a form
\begin{equation}
(\omega_0^2-\omega^2)u_i=\frac{QE_i}{M}-\frac{1}{M}\frac{\partial H_{int}}{\partial u_i}
\,,
\label{ph}\end{equation}
where $M$ is the reduced ion mass of the cell, $Q$ is the charge corresponding to the optical vibration, and $\omega_0$ is the frequency of the considered mode. Here, the last term represents the electron-phonon interaction. Using the Boltzmann equation (\ref{df}), we rewrite this term as follows
\begin{equation}
-\frac{1}{M}\frac{\partial H_{int}}{\partial u_i}=-\frac{u_i}{MN}\sum_{s}\int\frac{\omega\tau\zeta^2_i({\bf p})}{\omega\tau+i}
\left(-\frac{\partial f_0 }{\partial \varepsilon}\right)\frac{d^3{\bf p}}{(2\pi\hbar)^3}\,.
\label{ph1}\end{equation}
The term with electric field in the Boltzmann equation disappears in the integration over ${\bf p}$ due to the  velocity inversion ${\bf v }\rightarrow -{\bf v }$. The term with the wave vector ${\bf k}$ has to be omitted for the Raman phonon,
as the  vector ${\bf k}$ is determined in this case by the laser frequency $\omega_i$ and  the optical phonon frequency satisfies the condition
$\omega\gg \omega_iv/c$.

 \begin{figure}[h]
\resizebox{.4\textwidth}{!}{\includegraphics{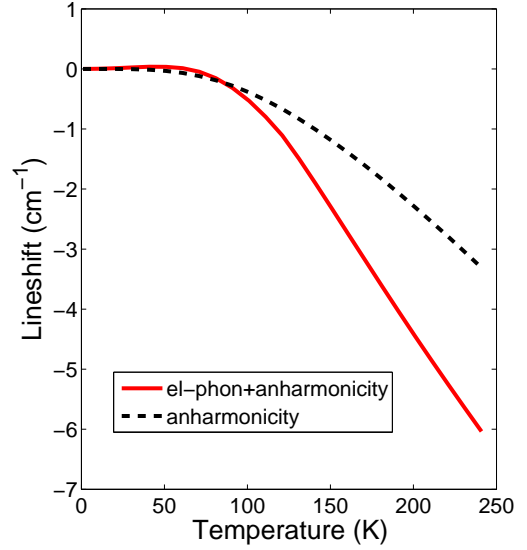}}
\caption{(Color online) Calculated shift of the Raman line $\omega=400 $ cm$^{-1}$ due to the electron-phonon interaction, and the lineshift in the Klemens channel (dashed line).}
\label{pos4}
\end{figure}

The electric field does not excited in the TO vibrations. Therefore, supposing $E=0$ and integrating over the energy $\varepsilon$ instead of $p$, we find from Eqs. (\ref{ph}) and (\ref{ph1}) the lineshift $\delta\omega$ and linewidth $\delta\Gamma$ determined by the electron-phonon interaction  as
\begin{equation}\delta\omega_{TO}-i\delta\Gamma_{TO}=\frac{1}{2MN}\sum_{s}\int\frac{\tau(\omega\tau-i)\zeta^2({\bf p})dS}{(\omega^2\tau^2+1)v(2\pi\hbar)^3}|_{\varepsilon=\varepsilon_ F}\,,
 \label{to}\end{equation}
 where $dS$ is an element of  the Fermi surface, $v$ is the Fermi velocity.
Estimating $S=4\pi p_F^2$, $\zeta({\bf p})\sim \varepsilon_0/a$ and $\varepsilon_0^2\sim \omega^2M/m$, where $\varepsilon_0\sim 3 eV$ is the typical electron energy in metals, we obtain
$$\delta\omega_{TO}-i\delta\Gamma_{TO}\sim \frac{ap_F \tau\omega^2_{TO} }{2\pi^2\hbar(\tau\omega_{TO}+i)}\,.$$   

The equations (\ref{ph}) and (\ref{ph1}) allow to express the phonon displacement $u$ in terms the electric field $E$ and to calculate the phonon contribution $ uNQ$ into the polarization. We find the total dielectric permittivity, adding the contributions $\varepsilon_{\infty}$ of the filled bands  
\begin{equation}
\begin{array}{c}
\varepsilon(\omega)={\displaystyle\varepsilon_{\infty}-\frac{4\pi e^2}{3\omega}\sum_{s}\int \frac{\tau vdS}{(\omega\tau+i)(2\pi \hbar)^3}}+\\{\displaystyle \frac{4\pi NQ^2}{M}\left[\omega_0^2-\omega^2+\frac{\omega\tau}{MN}\sum_{s}\int \frac{\zeta^2({\bf p})dS}{(\omega\tau+i)v(2\pi\hbar)^3}\right]^{-1}}.
\end{array}\label{eps}\end{equation}

\begin{figure}[]
\resizebox{.4\textwidth}{!}{\includegraphics{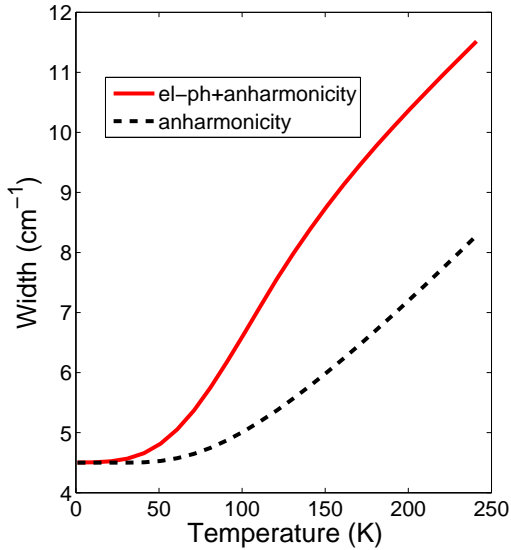}}
\caption{(Color online) Calculated temperature dependence of the width for the Raman $\omega=400 $ cm$^{-1}$  line at the ferromagnetic ordering, and the linewidth in the Klemens channel (dashed line).}
\label{width4}
\end{figure}

The frequency of the longitudinal phonon mode is determined by the condition $\varepsilon(\omega)=0$. In the absence of free carriers, one finds the frequency of the LO mode as follows
\[\omega_{LO}^2=\omega_0^2+\omega_{pi}^2,\]
where $\omega_{pi}^2=4\pi NQ^2/M\varepsilon_{\infty}$ is the ion plasma frequency squared.

Using Eq. (\ref{eps}), we find the LO frequency in the presence of carriers as
\begin{equation}\omega^2_{LO}-\omega_0^2=\frac{\omega}{(2\pi\hbar)^3MN}\sum_{s}\int \frac{\tau\zeta^2({\bf p})dS}{(\omega\tau+i)v}-\frac{\omega_{pi}^2}{\omega_{pe }^2}\omega(\omega+i\tau^{-1})\,,\label{lo}\end{equation}  
where the electron plasma frequencies squared  
\begin{equation} 
\omega_{pe }^2=\frac{4\pi e^2}{3\varepsilon_{\infty}}\sum_{s}\int \frac{vdS}{(2\pi \hbar)^3}\,\label{ie}\end{equation} 
is supposed to be large in comparison with $\omega_{pi}^2$.
We can put also $\omega=\omega_{LO}$ in the right-hand side of Eq. (\ref{lo}). Here the last term describes   the electric field screened by the free carriers. The main role plays the first term, which coincides 
 with the result for the TO mode, Eq. (\ref{to}), shown in Figs. \ref{pos4} and \ref{width4}, the results for the Klemens channel are taken from Ref. \cite{Fal}. 

We should emphasize that the temperature dependence of the linewidth and shift, Eq. (\ref{to}),  is determined mainly  by the electron collision rate $\tau^{-1}$ involving  also, for instance, in the dc conductivity. Thus, for a cubic crystal,  the dc conductivity, i.e. at $\omega=k=0$, writes
\[\sigma=\sum_{s}\frac{e^2}{3(2\pi \hbar)^3}
\int \tau vdS\,.\]
The  details of the electron density of states and of the deformation potential are responsible for peculiarities of the Raman line temperature dependence. 
\section{summary}
 
 The Klemens formula describes  the optical phonon width due to three-phonon anharmonic interactions. The corresponding lineshift matches with the linewidth.  In  such ferromagnets as CoS$_2$ with the low Curie temperature, these interactions are found to be too weak to describe quantitatively the experimental data and to explain the very large Raman linewidth and shift. Therefore, we propose the mechanism of the electron-phonon interaction attended with the effect of the ferromagnetic ordering on the electron bands. The deformation potential couples together  the  Boltzmann equation for electrons and the motion equation  for phonons producing the renormalization of the phonon frequency. The corresponding Raman line width and shift are  in  agreement with experiments in Ref. \cite{La}.  
 \section{acknowledgments}
 The author thank S. Lyapin and S. Stishov for information on their experiments prior the publication, and A. Varlamov for useful discussions. 
This work was supported by the Russian Foundation for Basic
Research (grant No. 13-02-00244A) and the SIMTECH Program, New Centure of Superconductivity: Ideas, Materials and Technologies (grant No. 246937).

\end{document}